\documentclass[useAMS,usenatbib,referee]{mn2e}
\usepackage{graphicx}
\usepackage{epsf}
\usepackage{epstopdf}

\title[%
Statistical distribution of current helicity]{%
Statistical distribution of current helicity in solar active regions
over the magnetic cycle. }
\author[Y.~Gao, T.~Sakurai, H.~Zhang, K.~Kuzanyan, D.~Sokoloff]
{Y.~Gao$^{1}$\thanks{Email: gy@bao.ac.cn},
T.~Sakurai$^2$\thanks{Email: sakurai@solar.mtk.nao.ac.jp},
H.~Zhang$^1$\thanks{Email: hzhang@bao.ac.cn},
K.M.~Kuzanyan$^{1,2,3}$\thanks{Email: kuzanyan@gmail.com},
D.~Sokoloff$^{1,4}$\thanks{Email: sokoloff.dd@gmail.com}\\
$^{1}${Key
Laboratory of Solar Activity, National Astronomical Observatories, Chinese Academy of Sciences, Beijing 100012, China}\\
$^2${National Astronomical Observatory of Japan, 2-21-1 Osawa, Mitaka, Tokyo 181-8588}\\
$^3${IZMIRAN, Troitsk, Moscow Region 142190, Russia}\\
$^{4}${Department of Physics, Moscow State University,
Moscow,119992, Russia}}

\begin{document}

\date{20120926}


\maketitle

\label{firstpage}

\begin{abstract}
The current helicity in solar active regions derived from vector
magnetograph observations for more than 20 years indicates the
so-called hemispheric sign rule; the helicity is predominantly
negative in the northern hemisphere and positive in the southern
hemisphere. In this paper we revisit this property and compare the
statistical distribution of current helicity with Gaussian
distribution using the method of normal probability paper. The data
sample comprises 6630 independent magnetograms obtained at Huairou
Solar Observing Station, China, over 1988-2005 which correspond to
983 solar active regions. We found the following. (1) For the most
of cases in time-hemisphere domains the distribution of helicity is
close to Gaussian. (2) At some domains (some years and hemispheres)
we can clearly observe significant departure of the distribution
from a single Gaussian, in the form of two- or multi-component
distribution. (3) For the most non-single-Gaussian parts of the
dataset we see co-existence of two or more components, one of which
(often predominant) has a mean value very close to zero, which does
not contribute much to the hemispheric sign rule. The other
component has relatively large value of helicity that often
determines agreement or disagreement with the hemispheric sign rule
in accord with the global structure of helicity reported by Zhang et
al. (2010).

\end{abstract}


\section{Introduction}

Recently there has been significant progress in collection and
interpretation of observational data on vector magnetic fields in
solar active regions which enables us to compute the values of
current helicity (a measure of departure of magnetic fields from
mirror symmetry) averaged over active regions (Seehafer 1990;
Pevtsov, Canfield and Metcalf 1995; Bao and Zhang 1998). The data
have been averaged over latitude and time in the solar cycle as well
(Zhang et al. 2010). The research so far has demonstrated important
properties of this quantity and its regular variation in the course
of the solar cycle. The helicity is generally negative in the
northern hemisphere and positive in the southern hemisphere; the
so-called hemispheric sign rule (HSR) for helicity. This rule,
however, may occasionally be violated in the activity minimum
periods (Hagino and Sakurai 2005; Zhang et al. 2010; Hao and Zhang
2012).

Helicity in the solar atmosphere has been noted as an important
agent which constrains magnetic field dissipation in the solar
corona. Current helicity plays an important role in the solar dynamo
theory as an observational proxy of the $\alpha-$effect as it
controls a dynamical back-reaction of the magnetic field to the
motion of the media which suppresses and stabilizes the generation
of the magnetic field (e.g. Kleeorin et al. 2003; Zhang et al.
2006). From a theoretical point of view the average helicity has the
meaning of a quantity averaged over an ensemble of turbulent
fluctuations in a small physical volume. This physical volume may
contain a limited number of convective cells, and so we may expect
it to vary significantly in space and time. The data on current
helicity are indeed very fluctuating within a given active region as
well as during its evolution (Zhang et al. 2002). Similar
fluctuations can be observed in the current helicity averaged over
latitude and time in the solar cycle.

However, the observation of current helicity is a complex process
which deals with initially imperfect data and involves highly
non-trivial reduction processes. These difficulties must be overcome
by collecting reliable datasets, because noise in the data will
degrade the reliability in the analysis (e.g. Abramenko et al. 1996;
Bao and Zhang 1998; Bao et al. 2001; Hagino and Sakurai 2004, 2005).
For better understanding of the reliability of the analysis of
current helicity previously undertaken, we hereby aim to study its
statistical distribution.

On the other hand the current helicity as a quantity responsible for
mirror asymmetry of solar magnetic field is expected to be
fluctuating also from a theoretical viewpoint: A usual expectation
is that the degree of mirror asymmetry in dynamo mechanisms will be
about 10\%. This practical estimate as originating from Parker
(1955) means that we have to isolate stable features of current
helicity distributions on a background of physical fluctuation which
may be ten times greater than the average value. This fluctuation is
not due to observational uncertainties and cannot be reduced by
improvement of observational techniques.

The expected substantial fluctuation in the current helicity data
has to be carefully taken into account in estimating the averaged
values of current helicity. Intrinsic (physical and true) dispersion
in the current helicity data around its mean value is, however,
interesting by itself.

The probability distribution of current helicity may be
substantially non-Gaussian. The point is that physical processes
responsible for the fluctuating nature of solar plasma can be
considered as an action of a product of independent evolutionary
operators rather than a sum of them resulting usually in a Gaussian
distribution. On the other hand, averaging over an active region
smoothes fluctuation and supports the Gaussian nature of
distribution.

Because of this, the statistical properties of current helicity
deserve to be addressed in full detail. Previously, the statistical
distribution of current helicity has been addressed only briefly as
a part of more general studies (e.g. Sokoloff et al. 2008).

In this paper we consider the distribution of available data over
the hemispheres of the Sun and their changes over solar cycles. We
compare this distribution with a Gaussian and separate the part of
the data which significantly deviates from a single Gaussian
distribution. Then, we consider the spatial and temporal properties
of this part in comparison with the other (Gaussian) part of the
data. We shall see that both parts exhibit similar properties and
behavior over the solar cycle.

\section{Observational data set}

This study is based on the data of photospheric vector magnetograms
of solar active regions obtained at Huairou Solar Observing Station,
China. The same database systematically covering 18 consecutive
years (1988-2005) was used in Zhang et al. (2010). The parameters
adopted there are $\alpha_{\rm av}$ (the average value of
$\alpha$;$\nabla \times {\bf B}  =\alpha {\bf B}$) and $H_{\rm c}$
(the integrated current helicity; $H_{\rm c} =\Sigma J_z B_z$). In
the present analysis we will use the same parameters.
(the integrated current helicity; $H_{\rm c} =\sigma J_z B_z$)
First, we analyze the entire database that comprises 6630
magnetograms of 983 different active regions (in terms of NOAA
region numbers). We are going to build our statistical analysis for
the whole bulk of available data. The majority of active regions are
represented by only one or very few magnetograms. However, there are
a few active regions that were recorded in 20 or more magnetograms.
In the next step we select one magnetogram for an active region.
Sometimes, and for some active regions that were observed within a
short time interval, the helicity parameters may be close with each
other. For these magnetograms, we select the one located nearest to
the center of the solar disk. Furthermore, we try to avoid choosing
the magnetograms of rapidly emerging active regions, except for
those regions which were recorded in only one magnetogram.
Nonetheless, this occurs rather rarely because such active regions
are usually large and well observed over several days.

\section{Method of statistical analysis}

We address the probability distribution of current helicity and its
deviation from Gaussian as it follows from observational data by two
statistical tools.

First of all, we divide the data into two hemispheres and produce
histograms of the current helicity distribution for certain time
intervals which contain 50 measurements. We then
approximate them by multiple Gaussians. In practice it appears that
a mixture of two Gaussians is sufficient to reproduce the histograms
with a reasonable accuracy.

This method, being very practical, may miss in principle substantial
deviations from Gaussian statistics which occurs with low
probability, i.e. intermittent features in the probability
distribution. A tool to address an intermittent feature is the
so-called Normal Probability Paper (NPP) test which is organized as
follows (see, e.g., Chernoff and Lieberman 1954).

Let our set contain $N$ active regions.  Let $n$ active regions have
current helicity density $\chi_{\rm c}$ lower than $x$. Then the
probability for $\chi_{\rm c}$ to be lower than $x$ is estimated as
$P = n/N$. Let $\xi$ be a Gaussian variable with the same mean value
$\mu$ and standard deviation $\sigma$ as $\chi_{\rm c}$ and let $y$
be the value for which the probability for $(\xi - \mu)/\sigma$ to
be lower than $y$ is $P$. The results for various $x$ values are
plotted in the $(x, y)$-plane and can be compared with the
cumulative distribution function (CDF) for a Gaussian distribution
which gives a straight line in this coordinate space. If the
observational data deviate substantially from this straight line, it
is an indication of a non-Gaussian nature of observational data (see
 e.g. Sokoloff et al. 2008 for details).

For illustration let us produce a set of random Gaussian variable
$xx_1$ ($N=1500$, $\mu_1=0.098$, and $\sigma_1=1.05$) by IDL
function ``randomn''. Then we produce another set of random Gaussian
variable $xx_2$ ($\mu_2=-0.28$ and $\sigma_2=0.42$). The GAUSS\_CVF
function computes the cutoff value $V$ in a standard Gaussian
distribution with a mean of 0.0 and a variance of 1.0 such that the
probability that a random variable $X$ is greater than $V$ is equal
to a user-supplied probability $P$ (see Figure 1). The upper panel
shows the probability distribution functions (PDFs) of two Gaussian
components and their sum (with equal weights). In the bottom panel
we show the relationship between the variable and the Gaussian CDF
with the same mean value and standard deviation for three kinds of
variables. The correlation for Gaussian variables is well fit by a
straight line but not for the sum of the components, as shown in
Figure 1 by the red curve. The slope of the straight line is
determined by the mean value, and the $y$-intercept is determined by
the standard deviation of the corresponding dataset.

\section{Results}

\begin{figure*}
\center
\includegraphics[angle=0,scale=1.2]{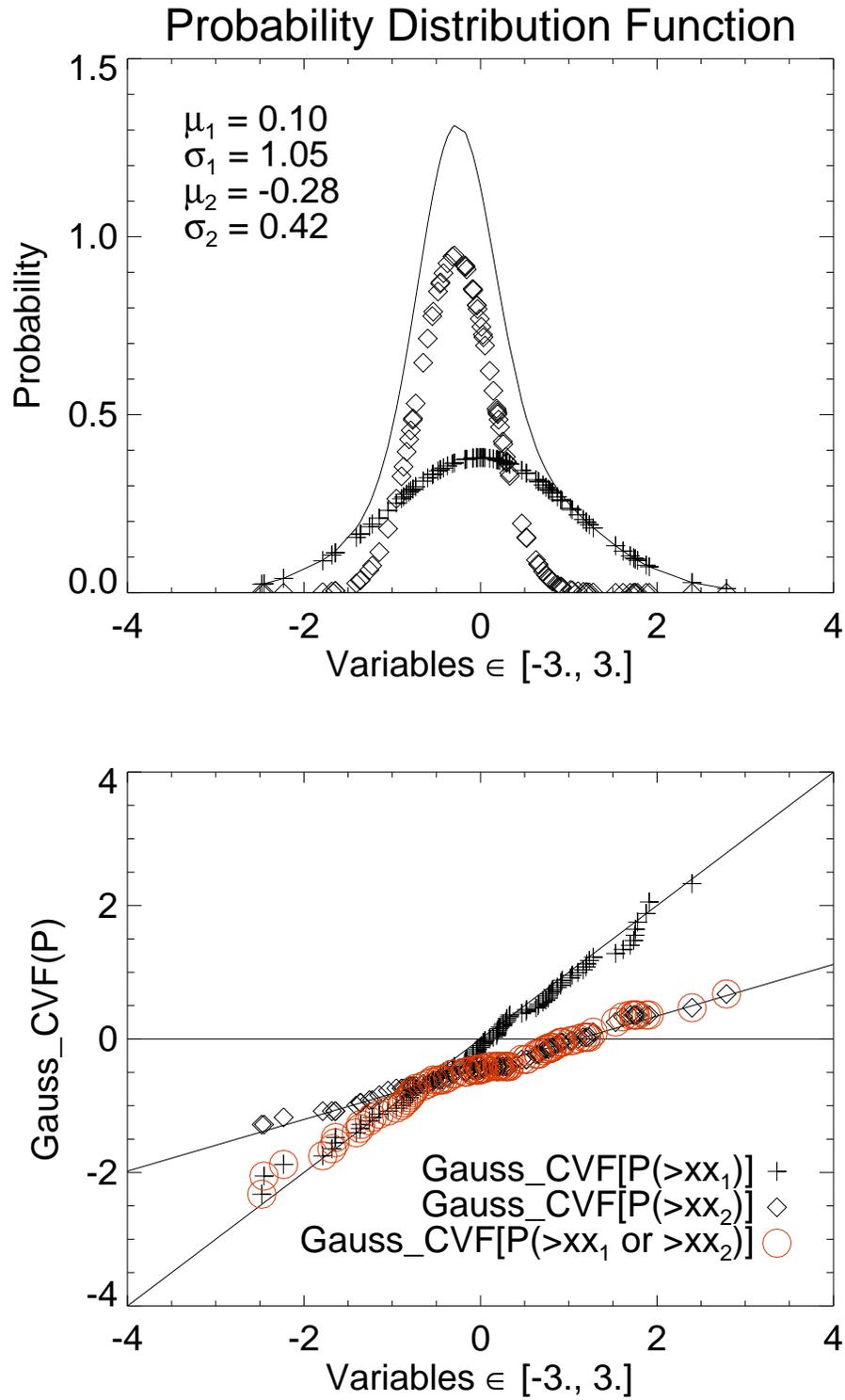}
\caption{Probability Distribution Function (PDF) of a set of random
Gaussian variable (upper panel) and correlation between the variable
and the Gaussian Cumulative Distribution Function (CDF) with the
same mean value and standard deviation (bottom panel).}
\end{figure*}

The entire sample includes 983 active regions, in which 464 (519) in
the northern (southern) hemisphere, respectively. Tables 1--4
present the detailed information about the Gaussian fitting to the
distribution function of the observed parameters. The columns
indicate; ``\#'': the sequence number of subgroups, ``Start''
(``End''): the earliest (last) measurement in the subgroup, $\delta
T$: the length of the time epoch between ``Start'' and ``End'',
$\mu_0$ and $\sigma_0$: the mean value and standard deviation of the
subgroup, and $\mu_1$, $\sigma_1$, $A_1$, $\mu_2$, $\sigma_2$ and
$A_2$ are the parameters defining the components of the sum of
Gaussian function in the form:
\begin{eqnarray}
 f(x)
 =\frac{A_1}{\sqrt{2\pi} \sigma_1} \exp \left[ -{\frac{(x-\mu_1)^2}{2\sigma_1^2}}\right] \\ \nonumber
+ \frac{A_2}{\sqrt{2\pi} \sigma_2} \exp \left[
-{\frac{(x-\mu_2)^2}{2\sigma_2^2}} \right] \,,
\end{eqnarray}
where the component with subscript ``1'' represents the component
with a greater amplitude; $A_1 \ge A_2$. The fitting is not made on
the PDF but is made for CDF; the function we use for the fitting is
actually the integral of Equation (1), namely the sum of two error
functions. The ``Error'' denotes the deviation of the observed
values from the fitting with the sum of two error functions. The
``No'' column records the number of data points in each subgroup.

\begin{table*}
\caption{%
Results of fitting to the data of $\alpha_{\rm av}$ for ten data
subgroups in the northern hemisphere. The fit error is standard
deviation of the fitting curve from the observed values. The values
of amplitudes $A_1$ and $A_2$ for cases when the second components
is significant ($A_2>A_1/2$) are underlined. }

\begin{tabular}{llllllllllllll}\hline
$\sharp$ & Start  & End &  $\delta$T      &    $\mu_0$ &$\sigma_0$
&     $\mu_1$   &  $\sigma_1$  &          $A_1$ &       $\mu_2$
&    $\sigma_2$    &            $A_2$      & Error & No\\ \hline
      1  &   Apr-16,1988 &   May-11,1990   &     755      &     -0.0088     &        0.0176   &    -0.0046     &      0.0154      &     0.8932      &    -0.0405   &        0.0061    &       0.1068    &       0.0948      &       50 \\
      2  &   May-20,1990 &   Jan-22,1992   &     612      &     -0.0119     &        0.0231   &    -0.0057     &      0.0160      &     \underline{0.6506}      &    -0.0226   &        0.0347    &       \underline{0.3494}    &       0.1356      &       50 \\
      3  &   Jan-25,1992 &   Dec-18,1993   &     693      &      0.0021     &        0.0227   &    -0.0031     &      0.0071      &     \underline{0.5360}      &     0.0094   &        0.0353    &       \underline{0.4640}    &       0.1364      &       50  \\
      4  &   Dec-26,1993 &   May-24,1998   &    1610      &     -0.0065     &        0.0169   &    -0.0044     &      0.0231      &     \underline{0.5651}      &    -0.0082   &        0.0088    &       \underline{0.4349}    &       0.1840      &       50 \\
      5  &   May-31,1998 &   Jul-22,1999   &     417      &     -0.0042     &        0.0161   &    -0.0034     &      0.0126      &     0.9609      &    -0.0843   &        0.0055    &       0.0391    &       0.1330      &       50   \\
      6  &   Jul-23,1999 &   Jun-11,2000   &     324      &     -0.0030     &        0.0128   &    -0.0026     &      0.0176      &    \underline{0.5978}      &    -0.0029   &        0.0058    &       \underline{0.4022}    &       0.1127      &       50    \\
      7  &   Jun-12,2000 &   Dec-21,2000   &     192      &     -0.0046     &        0.0140   &    -0.0016     &      0.0091      &     0.8807      &    -0.0402   &        0.0287    &       0.1193    &       0.1369      &       50   \\
      8  &   Dec-25,2000 &   Sep-02,2001   &     251      &      0.0005     &        0.0200   &     0.0035     &      0.0306      &     \underline{0.5017}      &    -0.0015   &        0.0071    &       \underline{0.4983}    &       0.1204      &       50   \\
      9  &   Sep-22,2001 &   Jun-28,2003   &     644      &     -0.0038     &        0.0177   &    -0.0045     &      0.0168      &     0.9779      &     0.2665   &        0.3073    &       0.0221    &       0.1293      &       50   \\
      10  &    Jul-5,2003 &   Dec-23,2005   &     902      &     -0.0028     &        0.0123   &    -0.0057     &      0.0211      &     \underline{0.5652}      &     0.0008   &        0.0035    &       \underline{0.4348}    &       0.2200      &       14  \\ \hline
\end{tabular}

\end{table*}

Table 1 shows the result of Gaussian fitting/decomposition for
$\alpha_{\rm av}$ in the northern hemisphere. We focus on the
relation between the sign of mean values of two Gaussian components
and the HSR. It is found that there are merely one $\mu_1$ (Row 8)
which violates the HSR from Dec-25, 2000 to Sep-02, 2001, while the
corresponding $\mu_2$ follows the HSR. In contrast, there are in
total three $\mu_2$'s (Rows 3, 9 and 10) which violate the HSR.
These epochs are ``Jan-25, 1992'' to ``Dec-18, 1993'', ``Sep-22,
2001'' to ``Jun-28, 2003'' and ``Jul-5, 2003'' to ``Dec-23, 2005'',
and their amplitudes $A_2$ are 0.46, 0.02 and 0.43, respectively. We
would like to stress here that the number of data points in the
latter subgroup is only 14, which is much less than in the other
subgroups. Therefore the fitting for this subgroup is not that
reliable. Nevertheless, the fitting in the third subgroup, i.e.,
from ``Jan-25, 1992'' to ``Dec-18, 1993'', is rather convincing. As
$A_2$ is 0.46 for the former group, both components are clearly seen
in the second row of Figure 2. In Figure 2 we also give fitting
results for the other three epochs. Their distributions are nicely
represented by two Gaussians and both components obey the HSR.

\begin{figure*}
\center
\includegraphics[angle=0,scale=1.2]{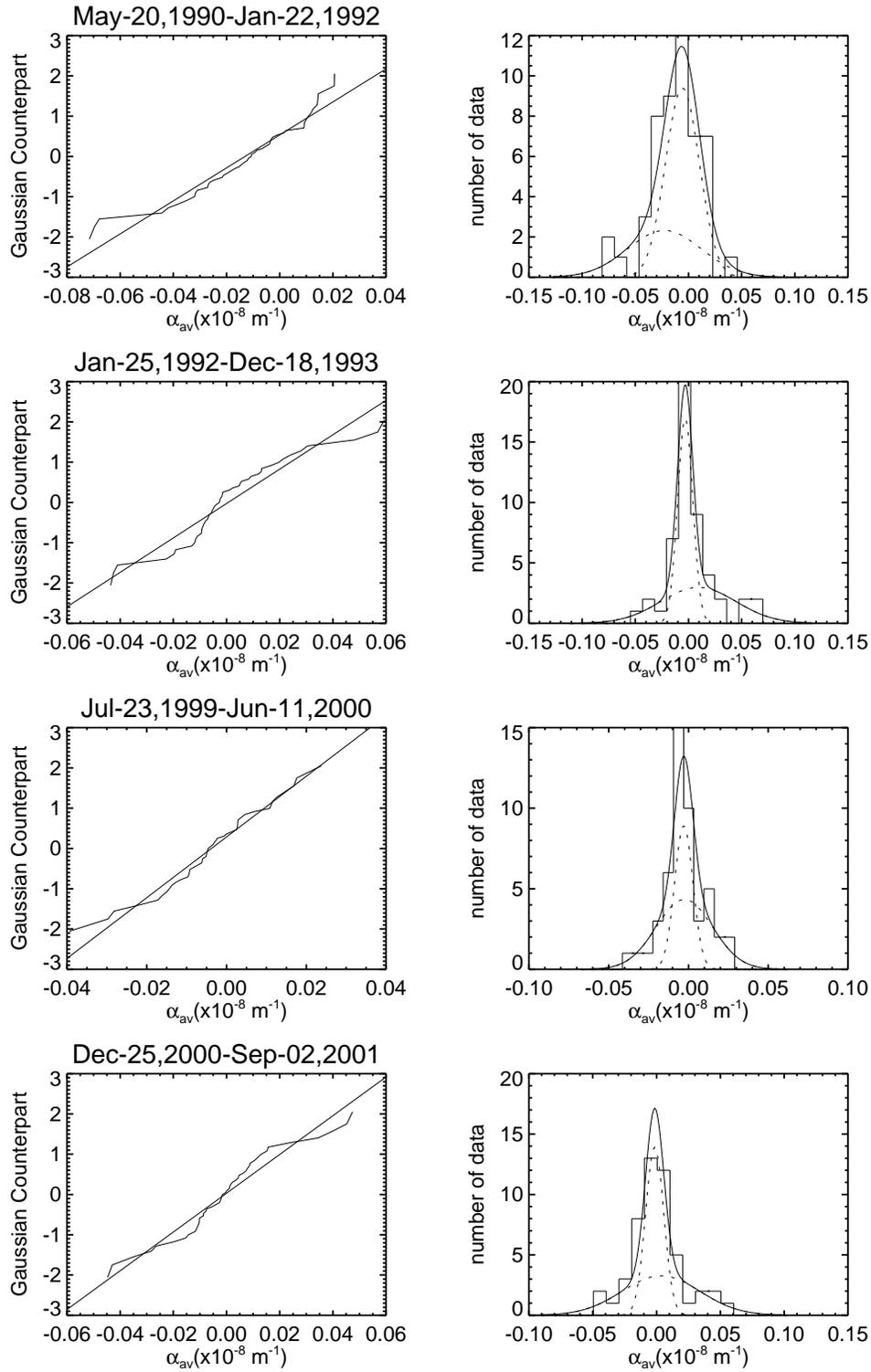}
\caption{Four cases of the distributions of $\alpha_{\rm av}$ in the
northern hemisphere. The left panels show NPP and are annotated by
the start and ending dates of the group. The right panels show the
data histogram and decomposed two Gaussians (dotted curves) and
their sum (solid curve).}
\end{figure*}

\begin{table*}
\caption{%
Results of fitting to the data of $\alpha_{\rm av}$ for eleven data
subgroups in the southern hemisphere. The fit error is standard
deviation of the fitting curve from the observed values.  The values
of amplitudes $A_1$ and $A_2$ for cases when the second component is
significant ($A_2>A_1/2$) are underlined. }
\begin{tabular}{llllllllllllll}\hline
$\sharp$ & Start  & End &  $\delta$T      &    $\mu_0$ &$\sigma_0$
&     $\mu_1$   &  $\sigma_1$  &          $A_1$ &       $\mu_2$
&    $\sigma_2$    &            $A_2$      & Error & No\\ \hline
       1  &    Apr-26,1988  &   Feb-25,1990    &    670    &      -0.0039      &     0.0235&    -0.0003   &        0.0181     &      0.8223     &      -0.0227     &       0.0487      &      0.1777     &       0.0993     &         50 \\
       2  &    Mar-08,1990  &   Aug-11,1991    &    521    &       0.0068      &     0.0254&     0.0027   &        0.0162     &      0.8035     &       0.0283     &       0.0542      &      0.1965     &       0.1364     &         50 \\
       3  &    Aug-12,1991  &   May-08,1992    &    270    &       0.0053      &     0.0165&     0.0020   &        0.0198     &      0.6668     &       0.0128     &       0.0096      &      0.3332     &       0.0807     &         50 \\
       4  &    Jun-16,1992  &   Jul-21,1993    &    400    &       0.0090      &     0.0206&     0.0008   &        0.0123     &      0.6766     &       0.0287     &       0.0266      &      0.3234     &       0.0931     &         50  \\
       5  &    Jul-29,1993  &   Aug-27,1996    &   1125    &       0.0038      &     0.0188&    -0.0019   &        0.0106     &      \underline{0.6147}     &       0.0147     &       0.0279      &      \underline{0.3853}     &       0.0895     &         50  \\
       6  &    Nov-29,1996  &   Apr-23,1999    &    875    &       0.0053      &     0.0155&     0.0053   &        0.0208     &      \underline{0.6437}     &       0.0063     &       0.0057      &      \underline{0.3563}     &       0.1044     &         50  \\
       7  &    May-19,1999  &   Apr-12,2000    &    329    &      -0.0010      &     0.0118&    -0.0001   &        0.0091     &      0.9575     &      -0.0587     &       0.0056      &      0.0425     &       0.1029     &         50  \\
       8  &    Apr-15,2000  &   Nov-20,2000    &    219    &       0.0048      &     0.0160&     0.0050   &        0.0110     &      0.7850     &       0.0044     &       0.0351      &      0.2150     &       0.0927     &         50  \\
       9  &    Nov-27,2000  &   Oct-11,2001    &    318    &       0.0022      &     0.0231&     0.0011   &        0.0122     &      \underline{0.6237}     &       0.0053     &       0.0399      &      \underline{0.3763}     &       0.1328     &         50  \\
       10 &    Oct-22,2001  &   Oct-27,2003    &    735    &       0.0002        &    0.0170   &   0.0017   &        0.0163     &      0.9807     &      -0.0694     &       0.0039      &      0.0193     &       0.0992     &         50 \\
       11 &    Oct-28,2003  &   Dec-16,2005    &    780    &      -0.0045        &    0.0166   &  -0.0070   &        0.0238     &      \underline{0.6532}     &       0.0010     &       0.0074      &      \underline{0.3468}     &       0.1287     &         19\\   \hline
\end{tabular}
\label{tab:example}
\end{table*}

\begin{figure*}
\center
\includegraphics[angle=0,scale=1.2]{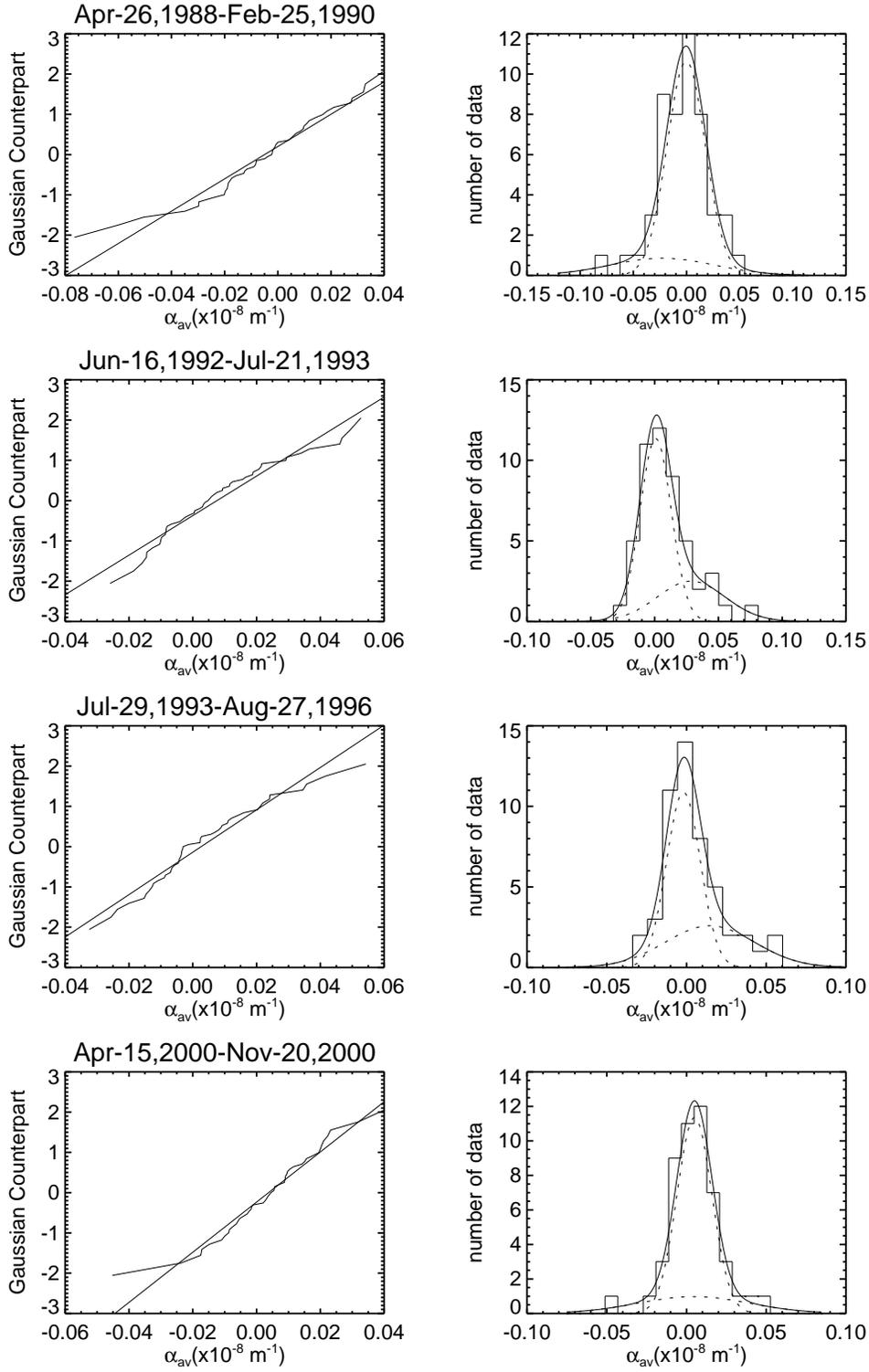}
\caption{Four cases of the distributions of $\alpha_{\rm av}$ in the
southern hemisphere.  The left panels NPP and are annotated by the
start and ending dates of the group. The right panels show the data
histogram and decomposed two Gaussians (dotted curves) and their sum
(solid curve).}
\end{figure*}

Table 2 shows the results of Gaussian fitting/decomposition for
$\alpha_{\rm av}$ in the southern hemisphere. It is found that four
$\mu_1$'s and three $\mu_2$'s violate the HSR. Among them, two cases
have both $\mu_1$  and $\mu_2$ violate the HSR. They occur in the
epochs ``Apr-26, 1988'' to ``Feb-25, 1990'', and ``May-19, 1999'' to
``Apr-12, 2000'', respectively. The amplitudes of $A_1$ are 0.82 and
0.96 and $A_2$ are 0.18 and 0.04, respectively. Other two $A_2$ that
violate HSR have amplitudes of 0.04 and 0.02, respectively. Some
examples are given in Figure 3. The first row shows apparent
violation of two components from the HSR. Other three cases show two
components clearly, too. However, all of these components obey the
HSR.

\begin{table*}
\caption{Results of fitting to the data of $H_{\rm c}$ for ten
subgroups in the northern hemisphere. }
\begin{tabular}{lllllllllllllll}\hline
$\sharp$ & Start  & End &  $\delta$T      &    $\mu_0$ &$\sigma_0$
&     $\mu_1$   &  $\sigma_1$  &          $A_1$ &       $\mu_2$
&    $\sigma_2$    &            $A_2$      & Error & No\\ \hline
      1    &  Apr-16,1988   & May-11,1990  &    755   &  -0.0421      &         0.0758   &       -0.0268     &      0.0741     &      0.8927      &      -0.1492    &       0.0198    &       0.1073   &    0.0860      &       50    \\
      2    &  May-20,1990   & Jan-22,1992  &    612   &  -0.0774      &         0.2541   &     -0.0347    &       0.0953    &       0.8894   &       -0.8641     &      0.5144     &      0.1106      &     0.0688      &       50    \\
      3    &  Jan-25,1992   & Dec-18,1993  &    693   &   0.0544      &         0.2714   &      0.0157    &       0.1149    &       0.9140   &        0.5670     &      0.9986     &      0.0860      &     0.1649      &       50     \\
      4    &  Dec-26,1993   & May-24,1998  &   1610   &  -0.0319      &         0.0814   &       -0.0268     &      0.1084     &      \underline{0.6583}      &     -0.0355    &       0.0132    &       \underline{0.3417}   &     0.2020      &       50    \\
      5    &  May-31,1998   & Jul-22,1999  &    417   &  -0.0187      &         0.0860   &     -0.0094    &       0.0668    &       0.9321   &       -0.3537     &      0.1570     &      0.0679      &     0.0822      &       50      \\
      6    &  Jul-23,1999   & Jun-11,2000  &    324   &  -0.0078      &         0.0651   &     -0.0050    &       0.0748    &       0.8629   &       -0.0139     &      0.0223     &      0.1371      &     0.1012      &       50       \\
      7    &  Jun-12,2000   & Dec-21,2000  &    192   &  -0.0267      &         0.0753   &       -0.0064     &      0.0476     &      0.7282      &     -0.0816    &       0.1299    &       0.2718   &     0.1168      &       50      \\
      8    &  Dec-25,2000   & Sep-02,2001  &    251   &  -0.0154      &         0.0677   &     -0.0043    &       0.0480    &       0.9682   &       -0.3474     &      0.0013     &      0.0318      &     0.1337      &       50      \\
      9    &  Sep-22,2001   & Jun-28,2003  &    644   &  -0.0228      &         0.1047   &     -0.0075    &       0.1297    &       0.7312   &       -0.0536     &      0.0287     &      0.2688      &     0.0991      &       50      \\
      10   &   Jul-5,2003   & Dec-23,2005  &    902   &  -0.0817      &         0.1614   &       -0.0030     &      0.0675     &      0.7956      &     -0.4294    &       0.1380    &       0.2044   &     0.0679      &       14       \\ \hline
\end{tabular}
\end{table*}

\begin{figure*}
\center
\includegraphics[angle=0,scale=1.2]{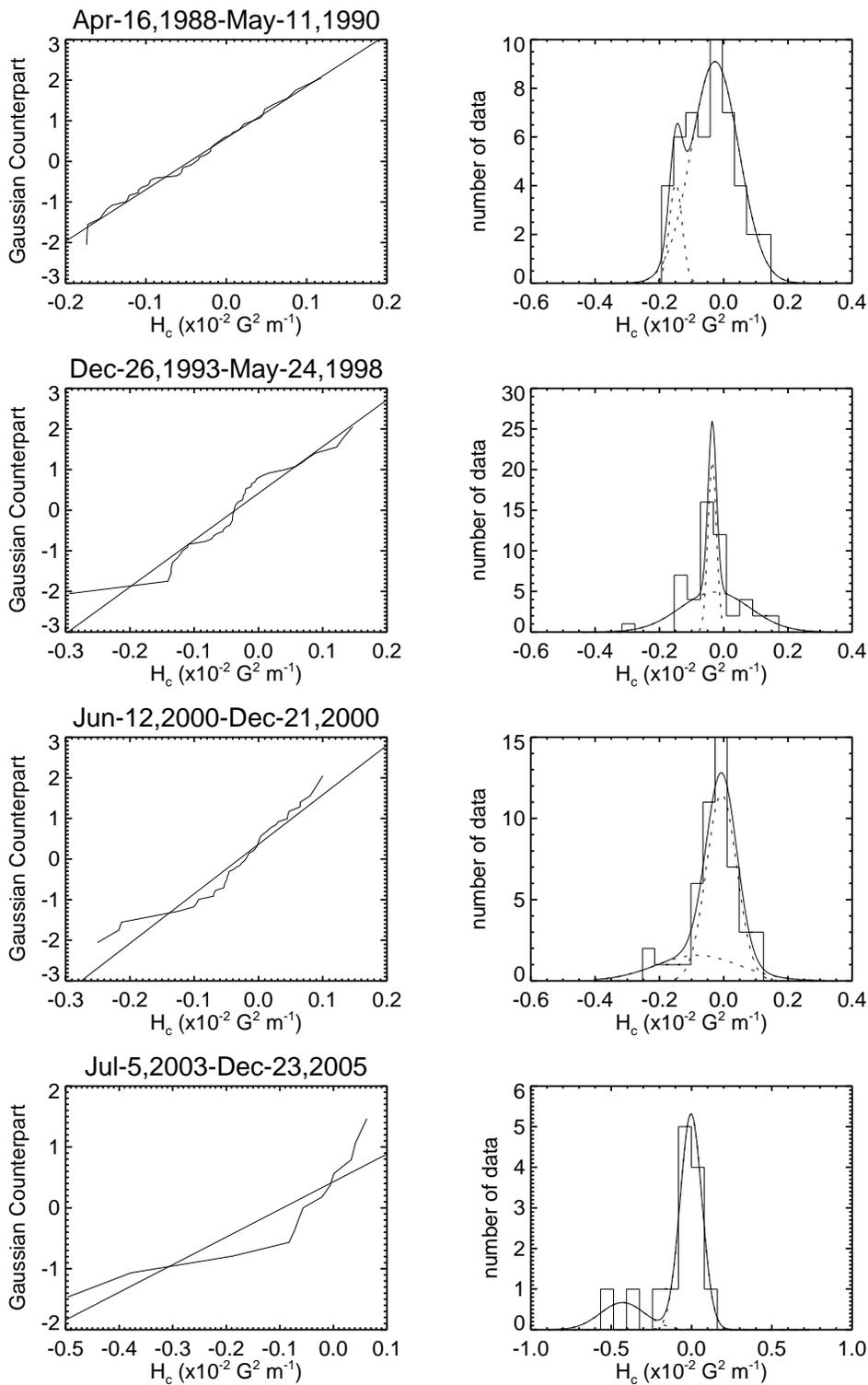}
\caption{Four cases of the distributions of $H_{\rm c}$ in the
northern hemisphere. The left panels are NPP and are annotated by
the start and ending dates of the group. The right panels show the
data histogram and decomposed two Gaussians (dotted curves) and
their sum (solid curves).}
\end{figure*}

We also perform comparative analysis for $H_{\rm c}$ in a similar
way. Table 3 shows the fitting results in the northern hemisphere.
For example, in  Row 3, i.e., from Jan-25, 1992 to Dec-18, 1993, the
means of both components  $\mu_1$ and $\mu_2$ show violation of the
HSR. Their amplitudes are 0.91 and 0.09. This is also consistent
with the results obtained for $\alpha_{\rm av}$. Some examples are
given in Figure 4.

\begin{table*}
\caption{%
Results of fitting to the data of $H_{\rm c}$ for eleven data
subgroups in the southern hemisphere. }
\begin{tabular}{llllllllllllll}\hline
$\sharp$ & Start  & End &  $\delta$T      &    $\mu_0$ &$\sigma_0$
&     $\mu_1$   &  $\sigma_1$  &          $A_1$ &       $\mu_2$
&    $\sigma_2$    &            $A_2$      & Error & No\\ \hline
       1 &   Apr-26,1988   & Feb-25,1990    &    670   &  -0.0273     &       0.1756     &       -0.0004     &      0.1008     &      0.9426     &     -0.8515     &      0.1035     &      0.0574   &        0.1190     &        50     \\
       2 &   Mar-08,1990   & Aug-11,1991    &    521   &   0.0061     &       0.2285     &        0.0622     &      0.0956     &      0.8886     &     -0.7620     &      0.3935     &      0.1114   &        0.1403     &        50     \\
       3 &   Aug-12,1991   & May-08,1992    &    270   &   0.0492     &       0.1122     &        0.0514     &      0.1234     &      0.9488     &      0.0628     &      0.0078     &      0.0512   &        0.0931     &        50      \\
       4 &   Jun-16,1992   & Jul-21,1993    &    400   &   0.1195     &       0.2721     &        0.0810     &      0.1102     &      0.9185     &      0.6108     &      1.0473     &      0.0815   &        0.2526     &        50     \\
       5 &   Jul-29,1993   & Aug-27,1996    &   1125   &   0.0176     &       0.0735     &        0.0115     &      0.0409     &      0.6867     &      0.0369     &      0.1377     &      0.3133   &        0.1160     &        50       \\
       6 &   Nov-29,1996   & Apr-23,1999    &    875   &   0.0331     &       0.0810     &        0.0130     &      0.0644     &      0.8737     &      0.1903     &      0.0312     &      0.1263   &        0.0989     &        50        \\
       7 &   May-19,1999   & Apr-12,2000    &    329   &   0.0053     &       0.0708     &       -0.0045     &      0.0386     &      0.6750     &      0.0317     &      0.1284     &      0.3250   &        0.0963     &        50       \\
       8 &   Apr-15,2000   & Nov-20,2000    &    219   &   0.0385     &       0.1084     &        0.0313     &      0.0547     &      0.6922     &      0.0617     &      0.2106     &      0.3078   &        0.1359     &        50       \\
       9 &   Nov-27,2000   & Oct-11,2001    &    318   &   0.0254     &       0.0807     &        0.0141     &      0.0438     &      0.8746     &      0.1172     &      0.2507     &      0.1254   &        0.1711     &        50       \\
       10&   Oct-22,2001   & Oct-27,2003    &    735   &   0.0198     &       0.1427     &        0.0213     &      0.0527     &      \underline{0.5970}     &      0.0260     &      0.2490     &      \underline{0.4030}   &        0.1354     &        50        \\
       11&   Oct-28,2003   & Dec-16,2005    &    780   &  -0.0133     &       0.1155     &       -0.0173     &      0.0949     &      0.8855     &     -0.4921     &      0.4071     &      0.1145   &        0.0873     &        19      \\ \hline
\end{tabular}
\end{table*}

\begin{figure*}
\center
\includegraphics[angle=0,scale=1.2]{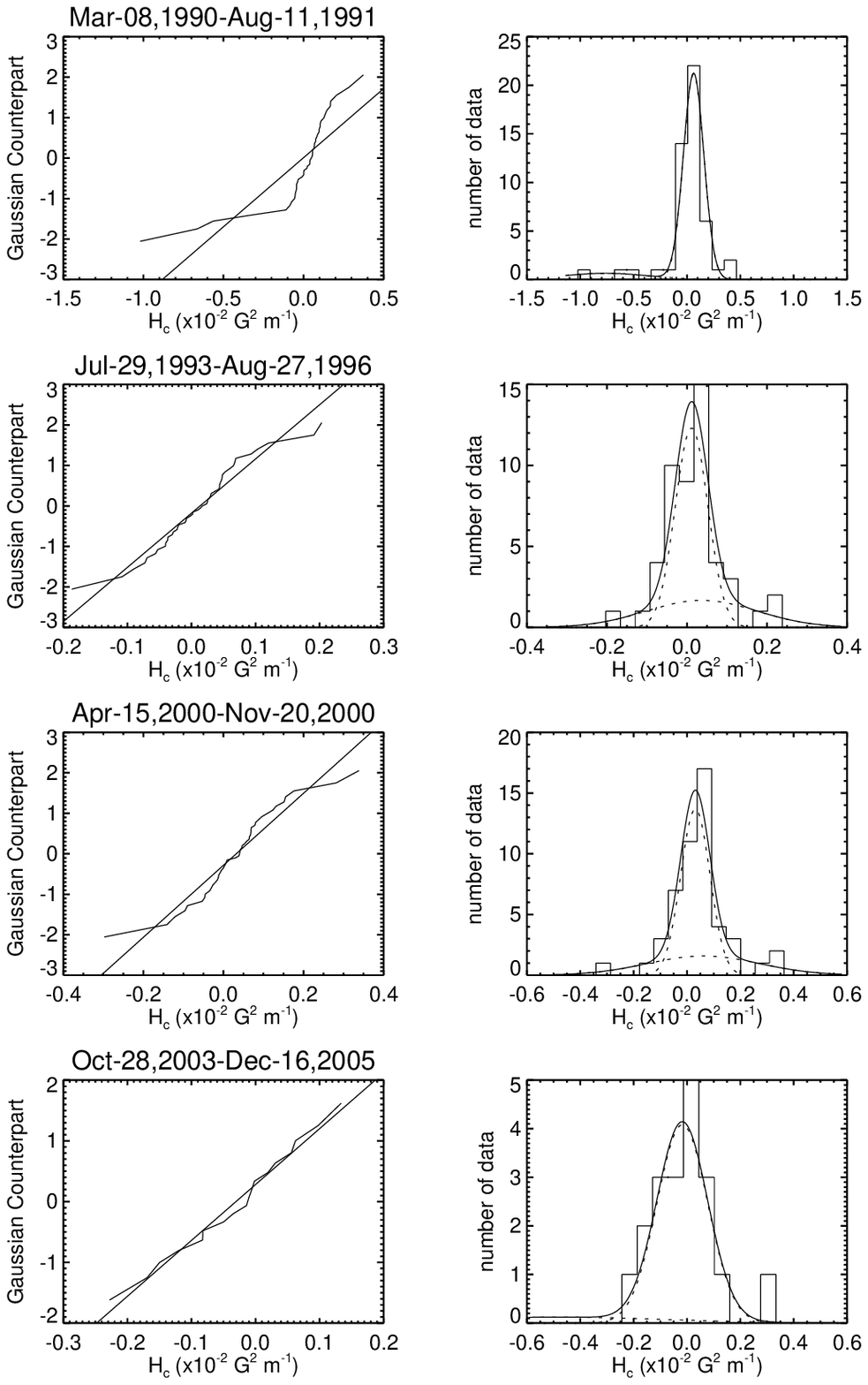}
\caption{Four cases of the distributions of $H_{\rm c}$ in the
southern hemisphere.  The left panels show NPP and are annotated by
the start and ending dates of the group. The right panels show the
data histogram and decomposed two Gaussians (dotted curves) and
their sum (solid curve).}
\end{figure*}

Table 4 shows the results of Gaussian fitting/decomposition for
$H_{\rm c}$ in the southern hemisphere. There are three $\mu_1$'s
that violate the HSR in Rows 1, 7 and 11, respectively. Their
amplitudes are 0.94, 0.68 and 0.89, respectively. In Rows 1 and 11,
the second components also violate the HSR, though with smaller
component amplitudes of 0.06 and 0.11. Another $\mu_2$ violating the
HSR occurs in the epoch of ``Mar-08, 1990'' to ``Aug-11, 1991'' (in
Row 2); its amplitude is 0.11. Some examples are given in Figure 5.

\begin{figure*}
\center
\includegraphics[angle=90,scale=0.6]{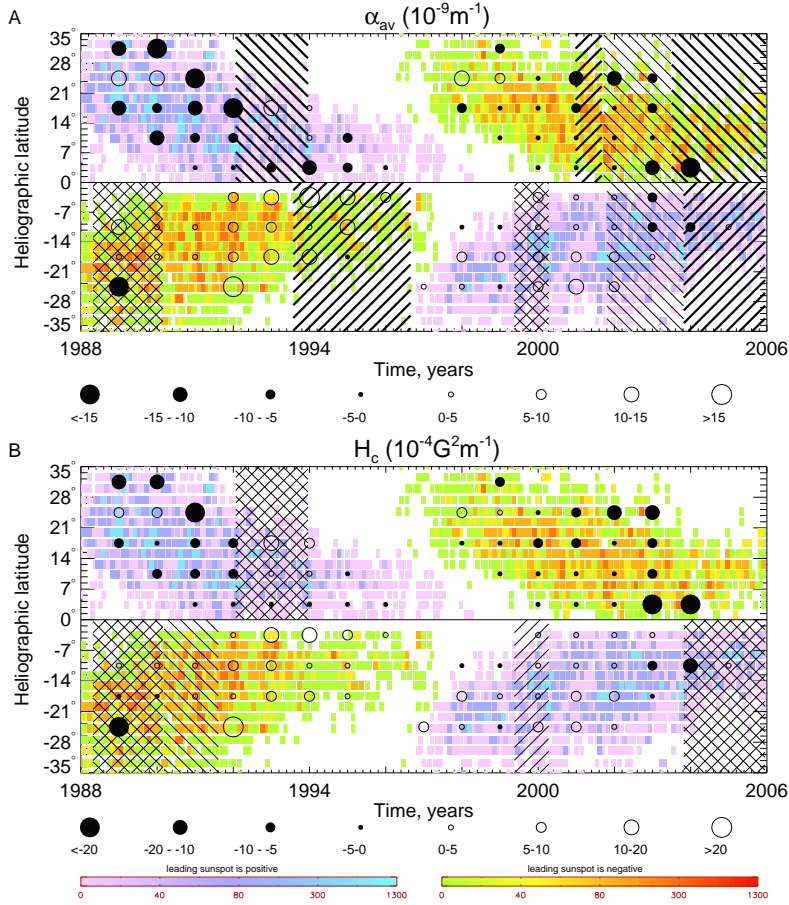}
\caption{Time-latitude distribution of Gaussian components for
selected 983 magnetograms of active regions (one magnetogram per
active region) over the period of 1988-2005. The background is the
butterfly diagram of current helicity plotted in time-latitude bins
as circles over the color plot of sunspot number density. Each bin
contains data coming from 7$^\circ$ in latitude and two year running
average in time. The 45$^{\circ}$ and $-45^{\circ}$ lines mark the
epoch in which the main component and sub-component violate the HSR,
respectively. The thick lines denote cases when the second component
is significant, i.e. the amplitudes satisfy the relation  $A_1 <
2A_2$. The upper and lower panels show the results for $\alpha_{\rm
av}$ and $H_{\rm c}$, respectively.}
\end{figure*}

To compare the epochs where the HSR is violated in the present
statistical analysis with the evolution and distribution of
$\alpha_{\rm av}$ and $H_{\rm c}$ obtained in our earlier papers
(Zhang et al. 2010), we mark these epochs with the inclined lines of
45$^{\circ}$ (the first component violates the HSR) and
$-45^{\circ}$ (the second component violates the HSR); see Figure 6.
Therefore, the crossed lines represent the cases when both the first
and the second components violate the HSR.

Here we estimate the uncertainty in determination of the two Gaussian mean values of the bi-modal distribution of the two parameters $\alpha_{\rm av}$ and
$H_{\rm c}$ under discussion. For that we use the 95 per cent Student's confidence intervals taking the standard deviations for each Gaussian component $\sigma_i$, where $i=1,2$, and computing the number of degrees of freedom as the overall number of available data points in each interval $n$ minus the number of fitting parameters, namely five. 
Then the expected errors in the mean values of the components would be
$\mu_i \pm {\sigma_i t_{n-5}}/{\sqrt{n-5}}$, where $t_{n-5}$ is Student's quantile for 95 per cent probability. 
We consider the violation of the rule significant if the error bars on the mean values that violate the HSR do not contain the zero value of the quantity under consideration.
The results are shown in Table 5-8. 
The mean values which violate the HSR are shown bold, and the cases of significant violation are underlined.
The cases for which violation of the HSR is statistically significant are shown in Fig. 7.

\begin{table*}
\caption{Statistical significance of HSR violation for $\alpha$ in the north hemisphere. Numbers in bold indicate violation of HSR, and they are underlined if regarded
statistically significant. }
\begin{tabular}{llllllllllllllll}\hline
$\sharp$ & Start & End &  $\mu_1$ & $\frac{\sigma_1 t_{n-5}}{\sqrt{n-5}}$  &  &  $\mu_2$&  $\frac{\sigma_2 t_{n-5}}{\sqrt{n-5}}$&  &\\ \hline
      3  & Jan-25,1992&Dec-18,1993&   -0.0031    &   0.0018       &    &        \underline{\textbf{0.0094}}   &  0.0088        &     \\
      8  &Dec-25,2000 &Sep-02,2001&   \textbf{0.0035}    &   0.0077       &     &       -0.0015   &  0.0018        &       \\
      9  &Sep-22,2001 &Jun-28,2003&   -0.0045    &    0.0042       &     &       \underline{\textbf{0.2665}}   &   0.0769        &       \\
      10  &Jul-5,2003 &Dec-23,2005&   -0.0057   &   0.0129       &       &       \textbf{0.0008}   & 0.0021        &      \\ \hline
\end{tabular}
\end{table*}

\begin{table*}
\caption{Statistical significance of HSR violation for $\alpha$ in the south hemisphere. Numbers in bold indicate violation of HSR, and they are underlined if regarded
statistically significant. }
\begin{tabular}{llllllllllllllll}\hline
$\sharp$ &Start&End&  $\mu_1$ & $\frac{\sigma_1 t_{n-5}}{\sqrt{n-5}}$  &   &  $\mu_2$&  $\frac{\sigma_2 t_{n-5}}{\sqrt{n-5}}$& &\\ \hline
       1  & Apr-26,1988&Feb-25,1990&\textbf{-0.0003}  &   0.0045    &     &       \underline{\textbf{-0.0227}}      &  0.0122    &   \\
       5  & Jul-29,1993&Aug-27,1996& \textbf{-0.0019}  &  0.0026    &      &         0.0147      &  0.0070    &      \\
       7  &May-19,1999 &Apr-12,2000& \textbf{-0.0001}  &  0.0023    &       &       \underline{\textbf{-0.0587}}      &  0.0014    &     \\
       10 &Oct-22,2001 &Oct-27,2003& 0.0017  &  0.0041   &    &       \underline{\textbf{-0.0694}}      &   0.0010   &   \\
       11 &Oct-28,2003&Dec-16,2005& \textbf{ -0.0070} &    0.0136   &     &        0.0010      &   0.0042   &    \\   \hline
\end{tabular}
\end{table*}

\begin{table*}
\caption{Statistical significance of HSR violation for $H_c$ in the north hemisphere. Numbers in bold indicate violation of HSR, and they are underlined if regarded statistically
 significant. }
\begin{tabular}{llllllllllllllll}\hline
$\sharp$ &Start&End&  $\mu_1$ & $\frac{\sigma_1 t_{n-5}}{\sqrt{n-5}}$  & &  $\mu_2$&  $\frac{\sigma_2 t_{n-5}}{\sqrt{n-5}}$&&\\ \hline
      3    &Jan-25,1992 &Dec-18,1993&     \textbf{0.0157}   &   0.0288            &       &    \underline{\textbf{0.5670}}    &    0.2500     &       \\ \hline               
\end{tabular}                                         
\end{table*}

\begin{table*}
\caption{Statistical significance of HSR violation for $H_c$ in the south hemisphere.  Numbers in bold indicate violation of HSR, and they are underlined if regarded statistically significant.}
\begin{tabular}{llllllllllllllll}\hline
$\sharp$ &Start&End&  $\mu_1$ & $\frac{\sigma_1 t_{n-5}}{\sqrt{n-5}}$  &   &  $\mu_2$&  $\frac{\sigma_2 t_{n-5}}{\sqrt{n-5}}$&  &\\ \hline
       1 &Apr-26,1988 &Feb-25,1990&      \textbf{-0.0004}     &     0.0252    &         &        \underline{\textbf{-0.8515}}   &    0.0259     &        \\
       7 &May-19,1999 &Apr-12,2000&      \textbf{-0.0045}     &    0.0097    &        &         0.0317   &    0.0322     &            \\
       11&Oct-28,2003&Dec-16,2005&       \textbf{-0.0173}     &     0.0544    &         &       \underline{\textbf{-0.4921}}   &     0.2334     &           \\ \hline
\end{tabular}
\end{table*}

\begin{figure*}
\center
\includegraphics[angle=90,scale=0.6]{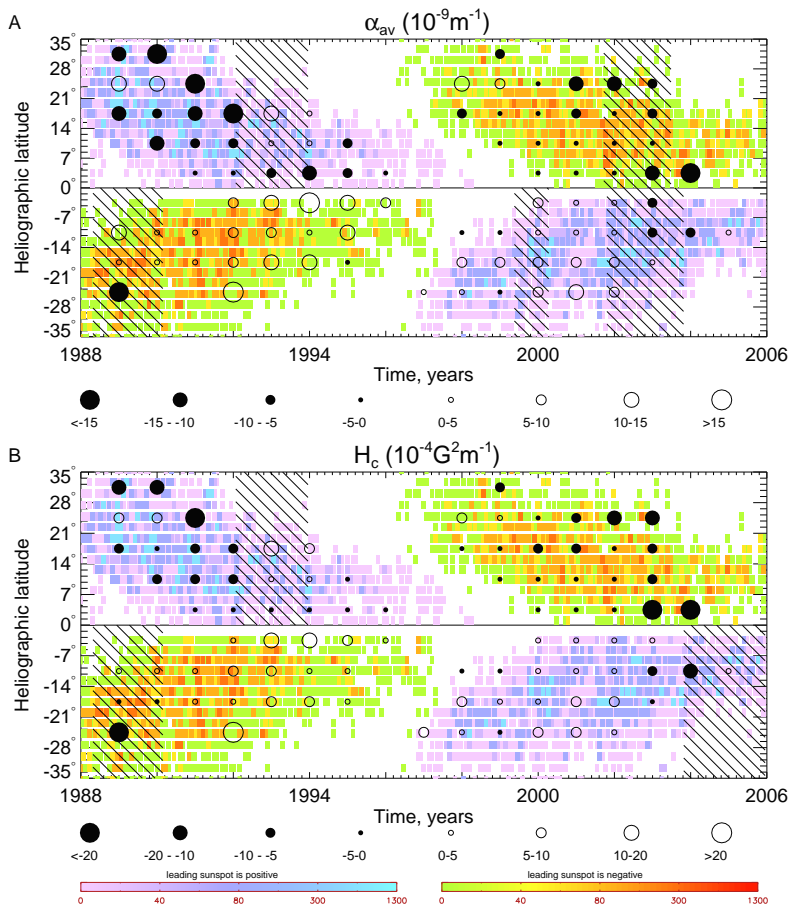}
\caption{Butterfly diagram with only those cases marked for which violation of the HSR is statistically significant. Notations are the same as in Fig. 6.}
\end{figure*}

The notable features common to two parameters $\alpha_{\rm av}$ and
$H_{\rm c}$ are as follows.
\begin{enumerate}
 \item  In the 22nd solar cycle there is an epoch of Apr-26,
 1988 - Feb-25, 1990 in the southern hemisphere, in which either the first
 or the second component of both parameters violated the HSR.
There is another epoch of Jan-25, 1992 --
 Dec-18, 1993 in the northern hemisphere, in which the second
 component of both parameters violated the HSR.

 \item
In the southern hemisphere of the 23rd solar cycle, the first
 component violated the HSR in the epochs of May-19, 1999 -- Apr-12,
 2000 and Oct-28, 2003 to Dec-16, 2005.

 \item
Looking at the cases of significant violation of the HSR in terms of Student's error bars we note that there are very few of them compared with the the overall cases of violation of the HSR by one or two Gaussian components of the bi-modal distribution.

  \item
We may also note that the cases of significant violation of the HSR occur not in the maximum of the solar cycle but in the phases of rise and fall.

\end{enumerate}
 The two parameters $\alpha_{\rm av}$ and $H_{\rm c}$ showed disparate results as follows.

\begin{enumerate}
 \item
For $\alpha_{\rm av}$ in the epochs of Dec-25, 2000 --
 Sep-02, 2001 in the northern hemisphere and Jul-29, 1993 -- Aug-27,
 1996 in the southern hemisphere, the first component violates the
 HSR. Also, in the epochs of Sep-22, 2001 -- Dec-23, 2005 in the northern
 hemisphere, the second component violates the HSR.
  These patches are not found for $H_{\rm c}$.

 \item
In the southern hemisphere, the first component of $\alpha_{av}$
 violates the HSR in the epoch of Jul-29, 1993 -- Aug-27, 1996,
 the second component
  of $\alpha_{av}$ violates the HSR in the epochs of May-19, 1999 -- Apr-12,
 2000 and Oct-22, 2001 -- Oct-27, 2003, but these are not
  seen for $H_{\rm c}$. In contrast, the second component of
 $\alpha_{\rm av}$ does not violate the HSR in the epochs of Mar-08, 1990
 -- Aug-11, 1991 and Oct-28,
 2003 -- Dec-16, 2005 but the second component of $H_{\rm c}$ does
 violate the HSR during those period.

  \item
The cases of significant violation of the HSR for both parameters 
$\alpha_{\rm av}$ and $H_{\rm c}$ coincide for cycle 22 but not 
for cycle 23. In cycle 23 there is only one case of significant violation 
of HSR for $H_{\rm c}$ but three cases for $\alpha_{\rm av}$. See Fig.~7 for details.

 \end{enumerate}

\section{Discussion}

We have investigated to what extent the current helicity and twist
data for solar active regions follow the Gaussian statistics and
what kind of message comes from the deviations from the Gaussian
distribution.

In our studies we have adopted the method of Normal Probability
Paper which has been developed for Gaussian distributions. Of
course, there was no reason to believe that the data must be ideally
distributed as a Gaussian. However, such analysis has shown relative
contributions of the sources of fluctuations as well as the
turbulent nature of the measured quantities.


Quite naturally, the statistics of current helicity and twist are
not exactly Gaussian. Here we confirmed the previous results of
Sokoloff et al. (2008). On the other hand, deviations from Gaussian
statistics are rather moderate and it looks often reasonable to
discuss the observed statistical distribution for given
space-latitude bins as a superposition of two Gaussian distributions
with specific means, standard deviations and amplitudes. We have not
encountered cases for which such fitting is impossible or
insufficient (but see the underlined values of $A_2$ in Tables
1--4). In other words, we have not detected traces of significant
intermittency in solar magnetic fields as its imprints in statistics
for the current helicity or twist. We appreciate that the
contemporary dynamo theory (see a review by Brandenburg, Sokoloff,
and Subramanian 2012) or analysis of the surface solar magnetic
tracers (Stenflo 2012) imply that a strong intermittency is
expected. Presumably, diffusive processes working during the rise of
magnetic flux tubes from the solar interior up to the surface might
have strongly smoothed the non-Gaussian features of the
distributions.

Formally speaking, the Gaussian distribution of the current helicity
was implicitly assumed when one estimates the error bars on the
current helicity averaged over a time-latitude bin (e.g., Zhang et
al. 2010). In practical respect, however, deviations from Gaussian
distribution obtained are small. Due to the Central Limit Theorem in
probability theory, one may expect only minor modifications to these
estimates and the non-Gaussian nature of distribution can be ignored
for this point.


We have isolated several epochs within the time interval covered by
observations where deviations from Gaussian statistics look
interesting and meaningful. We isolate the time bins with
substantial deviations from Gaussian distribution in two ways: when
the main Gaussian or the sub-component violates the HSR (Fig. 6), or
when the subcomponent is comparable with the main one.

Concerning the violation of the HSR (Fig. 6), we note a substantial
north-south asymmetry in the results: main part of the bins with HSR
violation belongs to the southern hemisphere. Remarkably, the
helicity data for the northern hemisphere for the 23rd cycle during
1997-2006 (with hemispheric averages) do not provide cases with HSR
violation at all.

We can summarize our main findings as follows.

\noindent 1. We have established that for the most of cases in
time-hemisphere domains the distribution of averaged helicity is
close to Gaussian.

\noindent 2. At the same time, at some domains (some years and
hemispheres) we can clearly observe significant departure of the
distribution from a single Gaussian, in the form of two- or
multi-component distributions. We are inclined to identify this fact
as a real physical property.


\noindent 3. For the most non-single-Gaussian parts of the dataset
we have established co-existence of two or more components, one of
which (often predominant) has a mean value very close to zero, which
does not contribute much to HSR. The other component has relatively
large value, whose sign is sometimes in agreement (for the data in
the maximum and shortly after the maximum of the solar cycle), or
disagreement (for example of 1989, just at the end of the rising
phase of cycle 22) with the HSR.

\noindent 4. Studies of the locations of the most
non-single-Gaussian parts over the time-latitude butterfly diagram
shows that these agreement and disagreement are in accord with the
global structure of helicity reported by Zhang et al. (2010, cf.
their Fig.2).

\noindent 5. We can interpret the result of multi-component
distribution of helicity in terms of the dynamo model which
addresses the origin of helicity in solar active regions. For
example, there may be spatial and time domains where the dynamo
mechanism does not work, or works differently.

We may note here that the agreement or disagreement with HSR at some
latitudes and times may be understood within the framework of solar
dynamo models (see, e.g. Zhang et al. 2012
and references therein). Discussion on the applicability of
particular dynamo models is beyond the framework of this paper and
will be addressed in our forthcoming studies.

\noindent 6. Another possible interpretation is that the active
regions which belong to multi-component distribution are
intrinsically different and formed at different depth or by
different mechanism.

\noindent 7. We may suggest that the formation of current helicity
in solar active regions may in general occur due to various physical
mechanisms at various scales. However, detailed investigation on
these mechanisms are yet to be done. This challenges both the dynamo
theory as well as the theory of flux tube/active region formation.

\section*{ACKNOWLEDGMENTS}
This work is partially supported by the National Natural Science
Foundation of China under the grants 11028307, 10921303, 11103037,
11173033, 41174153, 11178005, 11221063, by 
National Basic Research Program of China under the grant 2011CB811401 and 
by Chinese Academy of Sciences under
grant KJCX2-EW-T07 and XDA04060804-02. D.S. and K.K. would like to acknowledge support
from Visiting Professorship Programme of Chinese Academy or Sciences
2009J2-12 and thank NAOC of CAS for hospitality, as well as
acknowledge support from the NNSF-RFBR collaborative grant 13-02-91158 and 
RFBR under grants 12-02-00170 and 13-02-01183.. K.K. would like to 
appreciate Visiting Professorship
programme of National Observetories of Japan. We
thank the anonymous referee for his/her comments and suggestions
that helped to improve the quality of this paper.

\end{document}